# Charged impurity scattering in bilayer-graphene double layers

DANG KHANH LINH[1, 2]

[1]*Atomic Molecular and Optical Physics Research Group, Advanced Institute of Materials Science, Ton Duc Thang University, Ho Chi Minh City, Viet Nam (Email: dangkhanhlinh@tdtu.edu.vn )*
[2]*Faculty of Applied Sciences, Ton Duc Thang University, Ho Chi Minh City, Viet Nam.*

NGUYEN QUOC KHANH[3]
[3]*University of Science - VNUHCM, 227-Nguyen Van Cu Street, 5th District, Ho Chi Minh City, Viet Nam (Email: ngkhanh@hcmus.edu.vn).*

We consider a double-layer system made of two parallel bilayer graphene sheets separated by a dielectric medium. We calculate the finite-temperature electrical conductivity of the first layer due to charged impurities located in two layers. We study the effects of temperature, interlayer distance, dielectric constants and impurity concentration, carrier concentration on the electrical conductivity. We show the importance of charged impurities located in layer II in determining electrical conductivity of the first layer for small interlayer distance. The results in this paper give us more understanding about the long-range charged impurity scattering in bilayer graphene under effect of the second one.

## 1. Introduction

Since graphene has been discovered in 2004, transport properties of graphene and graphene-based structures have attracted an attention of many researchers. A lot of papers on the electrical conductivity, thermopower and thermal conductivity of monolayer [1-12], bilayer graphene (MLG, BLG) [13-19] have been published. Besides, properties of double-layer structures such as MLG-MLG [19-22], MLG-GaAs [23-26], BLG-GaAs [27-28], BLG-MLG [29-30] and BLG-BLG [31-33] have been also studied recently.

Hosono and Wakabayashi have calculated the carrier mobility of a MLG-MLG heterostructure at $T$ = 0K [34], while Parhizgar and Asgari [21] have considered similar systems and calculated the zero-temperature longitudinal resistivity and magnetoresistance of first graphene layer under effect of the second one. The authors of this paper have extended the work of Parhizgar and Asgari to finite-temperature case for BLG-BLG [35] and BLG-2DEG [36] double layers assuming that charged impurities are located only in the first layer. In this paper, we generalize the results of [35] to include the effects of impurities in both layers. We show that electrical conductivity of BLG (the first layer) decrease (i.e. resistivity of BLG increases) in presence of the charged impurities in the second layer. This is an interesting result that compares with some calculations of N. M. R. Peres et al for double layer graphene system [20]. Besides, we show that there is a nonmonotonicity of the electrical conductivity of BLG as function of $\varepsilon_2$ is observed at low temperatures and the substrate strongly reduces the Coulombic scattering due to charged impurities located in both layers. We show that the transition temperature is independent of $d, n_{i2}, \varepsilon_3, n_2$, and it sensitive to $\varepsilon_2$. Finally, we find that conductivity decreases with increasing carrier concentration $n_2$ for low temperatures and then remains almost constant.

## 2. Theory

We consider the double-layer BLG-BLG (hereafter referred to as 2BLG) system, shown in Fig. 1, consisting of a doped BLG ( layer 1 ) placed above another doped BLG ( layer 2 ) deposited on an insulating substrate. Two layers have equal carrier density ($n_1 = n_2 = n$) and are electrically isolated via an insulating spacer of thickness $d$ so that the tunneling of electrons between two layers can be neglected.

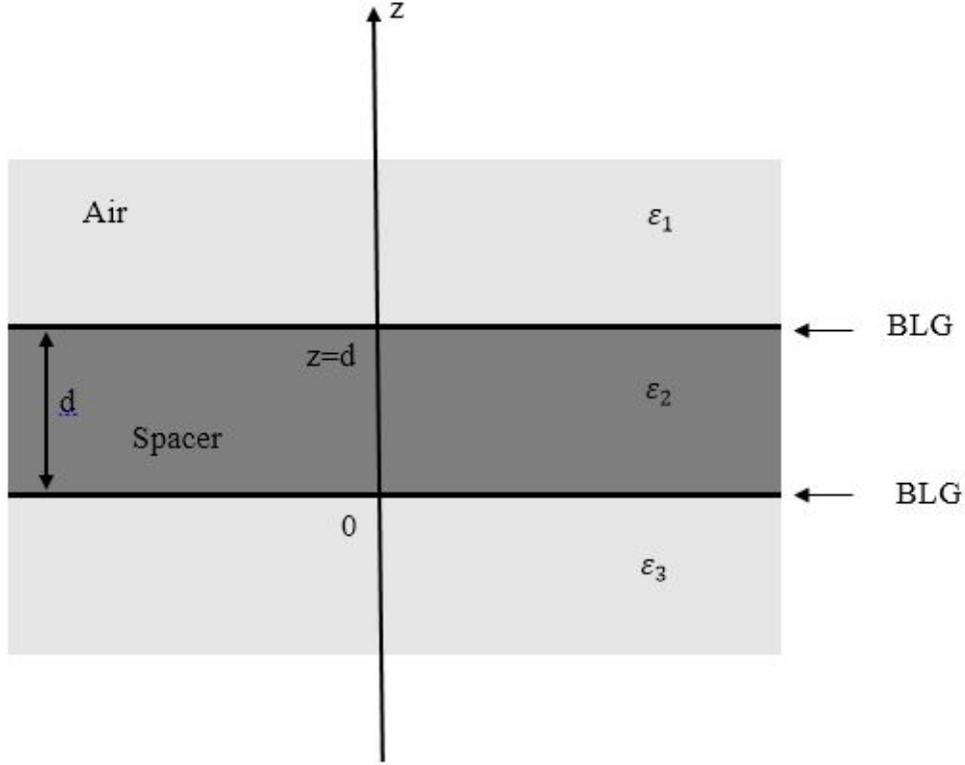

Fig. 1. A bilayer graphene double-layer system immersed in a three layered dielectric medium with the background dielectric constants $\varepsilon_1 = \varepsilon_{Air} = 1$, $\varepsilon_2$ and $\varepsilon_3$.

We assume that the charged impurities are located both in the first and second layer with concentration $n_{i,1}$ and $n_{i,2}$, respectively. To calculate the conductivity of layer 1 in presence of layer 2, we use the Boltzmann approach which is based on the relaxation time approximation and consider only charged impurity scattering. In this approach, the conductivity is given by [13, 37-39]

$$\sigma = \frac{N_0 e^2}{m^*} \int_0^\infty dE \left(-\frac{\partial f}{\partial E}\right) \tau(E,T) E \tag{1}$$

where $N_0 = \frac{2m^*}{\pi \hbar^2}$ is density of states in BLG, with $m^*$ being the effective mass of electron. $f(E_{\vec{k}}) = \frac{1}{e^{\frac{E-\mu}{k_B T}}+1}$ is the Fermi – Dirac distribution function, $\mu = E_F = \frac{\hbar^2 k_F^2}{2m^*}$ is the BLG chemical potential, $T_F = \frac{E_F}{k_B}$ is the BLG Fermi temperature and $\tau(E,T)$ is the total relaxation time of layer 1 given by [34]

$$\frac{1}{\tau_1} = \frac{1}{\tau_{11}} + \frac{1}{\tau_{12}} \tag{2}$$

where

$$\frac{1}{\tau_{1j}(E_{\vec{k}},T)} = \frac{n_{i,j}}{2\pi E_{\vec{k}}} \int_0^{2k} dq \frac{q^2[1 - 2(\frac{q}{2k})^2]^2}{\sqrt{4k^2 - q^2}} |W_{1j}(q,T)|^2. \tag{3}$$

Here $E_F$ ($k_F$) is the Fermi energy (Fermi wave-vector) of BLG and $W_{1j}(q,T)$ represents the effective interaction between electrons in layer 1 and impurities in layer $j = 1, 2$ [21, 34],

$$W_{11}(q,T) = \frac{V_{11}(q) + (V_{12}(q)^2 - V_{11}(q)V_{22}(q))\Pi_2(q,T)}{\varepsilon(q,T)} \tag{4}$$

$$W_{12}(q,T) = \frac{V_{12}(q)}{\varepsilon(q,T)} \tag{5}$$

where $V_{ll}(\vec{q})$ and $V_{12}(\vec{q})$ are the intra- and inter-layer electron – electron Coulomb interactions given by [21, 34]

$$V_{11}(q) = \left(\frac{4\pi e^2}{q}\right)\left(\frac{\varepsilon_2 + \varepsilon_3\tanh(qd)}{\varepsilon_2(\varepsilon_1 + \varepsilon_3) + (\varepsilon_2^2 + \varepsilon_1\varepsilon_3)\tanh(qd)}\right), \quad (6)$$

$$V_{22}(q) = \left(\frac{4\pi e^2}{q}\right)\left(\frac{\varepsilon_2 + \varepsilon_1\tanh(qd)}{\varepsilon_2(\varepsilon_1 + \varepsilon_3) + (\varepsilon_2^2 + \varepsilon_1\varepsilon_3)\tanh(qd)}\right), \quad (7)$$

$$V_{12}(q) = V_{21}(q) = \left(\frac{4\pi e^2}{q}\right)\left(\frac{\frac{\varepsilon_2}{\cosh qd}}{\varepsilon_2(\varepsilon_1 + \varepsilon_3) + (\varepsilon_2^2 + \varepsilon_1\varepsilon_3)\tanh(qd)}\right) \quad (8)$$

and $\varepsilon(q,T)$ is the RPA dielectric function of 2BLG,

$$\varepsilon(q,T) = \left(1 - \Pi_1(q,T)V_{11}(q)\right)\left(1 - \Pi_2(q,T)V_{22}(q)\right) - V_{12}(q)^2\Pi_1(q,T)\Pi_2(q,T) \quad (9)$$

with $\Pi_1(q,T)$ ($\Pi_2(q,T)$) being the finite-temperature static polarizability of the layer 1 ( layer 2) [13],

$$\Pi_1(q,T) = \Pi_2(q,T) = \Pi(q,T)$$
$$= -N_0 \int_0^\infty \frac{dk}{k^3}\left\{\sqrt{4k^4 + q^4} - k^2 - |k^2 - q^2|\right.$$
$$\left. + [f(E_{\vec{k}}) + f(E_{\vec{k}+2\mu})]\left[2k^2 - \sqrt{4k^4 + q^4} + \frac{(2k^2 - q^2)^2}{q\sqrt{q^2 - 4k^2}}\theta(q - 2k)\right]\right\} \quad (10)$$

### 3. Numerical results

#### *3.1. The dependence of conductivity of the first layer on the charged impurities located in layer II*

In this section we present numerical results for electrical conductivity using following system parameters: $\varepsilon_1 = \varepsilon_{\text{Air}} = 1$ and $m^* \approx 0.033 m_e$ where $m_e$ is the vacuum mass of electron [13, 40]. When not otherwise stated, the layer 2 is assumed to be placed on an $Al_2O_3$ substrate with $\varepsilon_3 = \varepsilon_{Al_2O_3} = 12.53$. We choose $n = 2 \times 10^{12}$ cm$^{-2}$ [14, 31] and $n_{i,1} = 1 \times 10^{11}$ cm$^{-2}$ [14], $n_{i,2} = 1 \times 10^{11} - 1 \times 10^{12}$ cm$^{-2}$ [14, 42].

In Fig. 2 we plot the finite-temperature electrical conductivity of the first layer in presence of the second one due to the screened Coulomb scattering. For a comparison, we show in the insets corresponding results for the case when the charged impurities are located only in the layer 1. It is obvious that the shape of electrical conductivity curves is similar in two cases with and without the impurities in the second layer. When the interlayer spacing is small, $d = 1nm$, $\sigma$ decreases with increase in the impurity concentration in the layer 2, $n_{i,2}$ and the decrease of $\sigma$ becomes remarkable for $n_{i,2} \geq 7n_{i,1}$ (ref. Fig. 3). This behavior show that the charged impurities located in layer II contribute significantly to determining bilayer electrical conductivity for small interlayer spacing $d = 1nm$. For large interlayer spacing, $d \geq 4\ nm$ (ref. Fig. 4), the independent on $n_{i,2}$ of $\sigma$ establishes that the effect of charge impurities in layer I is crucial. Finally, the transition temperature is found to be equal to $0.2T_F$ for all $d, n_{i2}$.

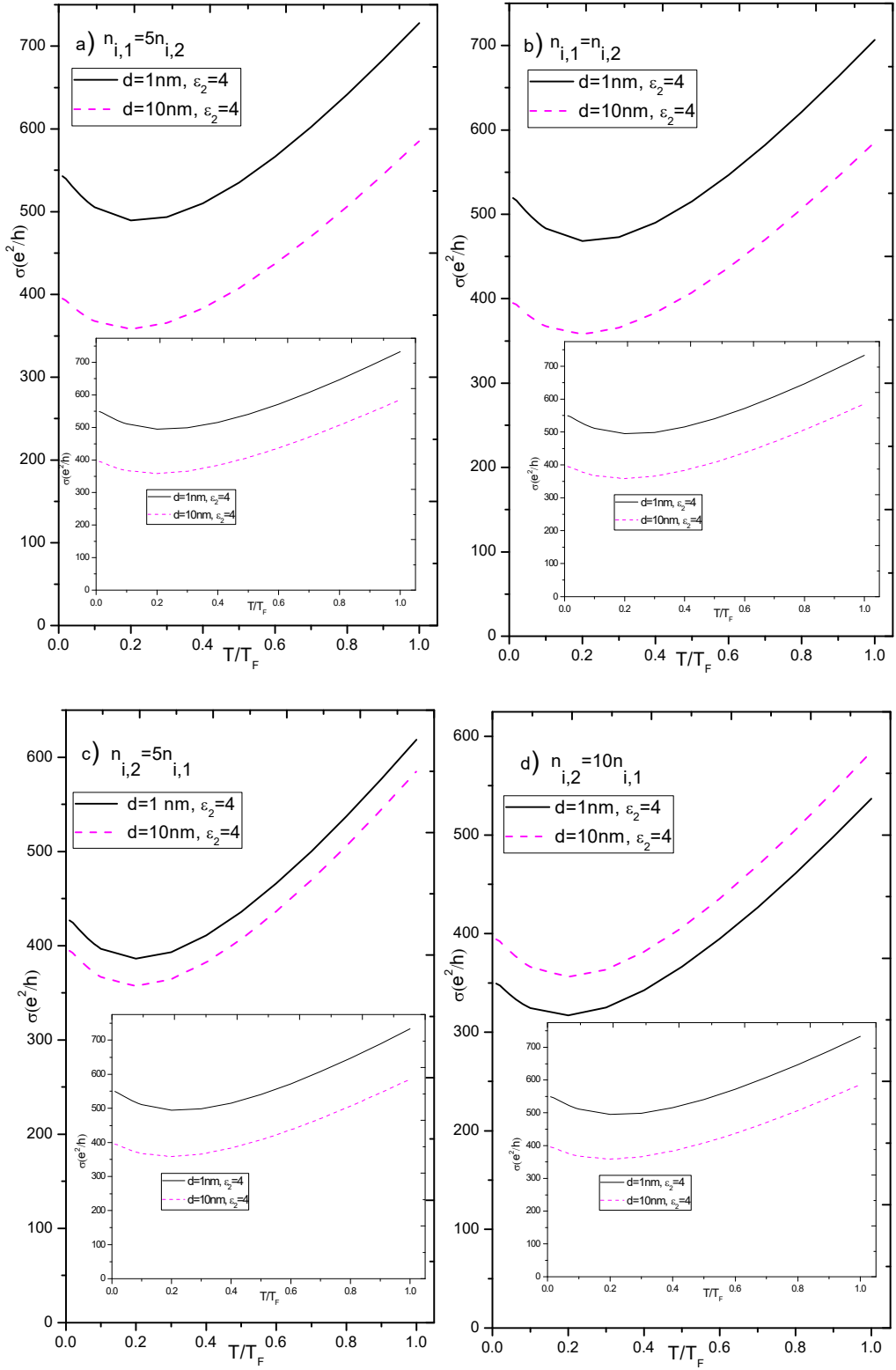

Fig. 2. The electrical conductivity $\sigma$ of layer 1 for 2BLG with $\varepsilon_2 = 4$, $\varepsilon_3 = 12.53$ as a function of $\frac{T}{T_F}$ for $d = 10nm$, $d = 1nm$, and $n_{i,1} = 5n_{i,2}$ (a), $n_{i,1} = n_{i,2}$ (b), $n_{i,2} = 5n_{i,1}$ (c), $n_{i,2} = 10n_{i,1}$ (d). The insets show corresponding results for the case when the impurities are absent in layer 2, $n_{i,2} = 0$.

In Fig. 3, we show the electrical conductivity $\sigma$ for 2BLG with $\varepsilon_2 = 4$, $\varepsilon_3 = 12.53$, $d = 1nm, d = 10nm$ as a function of impurity ratio $\frac{n_{i,2}}{n_{i,1}}$ for $T = 0.01T_F$, $T = 0.5T_F$ and $T = 0.9T_F$. The figure indicates that for small interlayer distances, $d = 1nm$, the conductivity decreases rapidly with increasing $n_{i,2}$ for all temperatures considered as expected. When $n_{i,2} \geq 7n_{i,1}$, $\sigma$ (at $d = 1nm$) is larger than $\sigma$ (at $d = 10nm$).

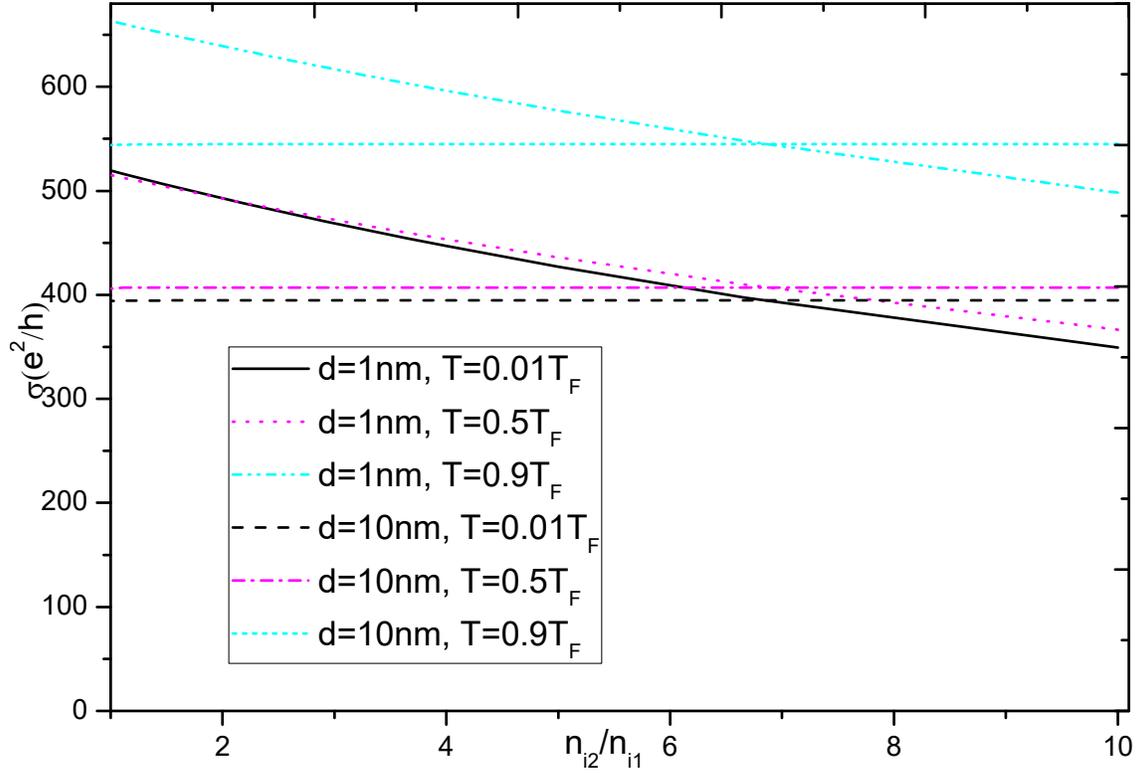

Fig. 3. The electrical conductivity of layer 1 for 2BLG with $\varepsilon_2 = 4$, $\varepsilon_3 = 12.53$ as a function of $\frac{n_{i,2}}{n_{i,1}}$ for $T = 0.01T_F$, $T = 0.5T_F$ and $T = 0.9T_F$.

### 3.2. *The dependence of conductivity of the first layer on spacer thickness*

To understand the effect of spacer thickness $d$, we show in Fig. 4 the electrical conductivity $\sigma$ as a function of $d$ for 2BLG with $\varepsilon_2 = 4$, $\varepsilon_3 = 12.53$ and $n_{i,2} = 10n_{i,1}$ in three cases $T = 0.01T_F$, $T = 0.5T_F$ and $T = 0.9T_F$. It is seen that the conductivity increases with $d$ for $d < 4\ nm$. For $d > 4\ nm$, $\sigma$ remains almost constant for all considered values of temperature.

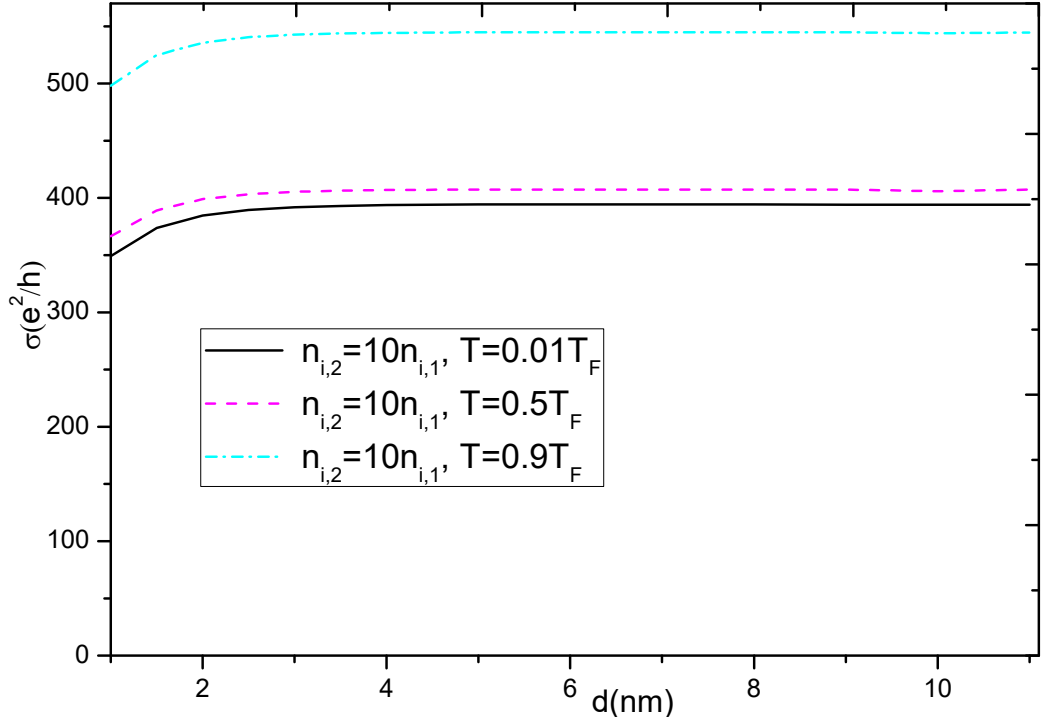

Fig. 4. The electrical conductivity of layer 1 of 2BLG with $\varepsilon_2 = 4$, $\varepsilon_3 = 12.53$, $n_{i,2} = 10n_{i,1}$ as a function of $d$ for $T = 0.01T_F$, $T = 0.5T_F$ and $T = 0.9T_F$.

### 3.3. *The dependence of conductivity of the first layer on dielectric constant of spacer*

To study the effect of spacer, we depict in Fig. 5 the dependence of electrical conductivity on temperature for 2BLG with $\varepsilon_3 = 12.53, d = 1nm, n_{i,2} = 10n_{i,1}$ for three different values of $\varepsilon_2$ and in Fig. 6 the conductivity with $\varepsilon_3 = 12.53, d = 1\ nm$ and $n_{i,2} = 10n_{i,1}$ as a function of dielectric constant $\varepsilon_2$ for three different values of temperature $T$. The results in Fig. 5 show that the conductivity increases with increasing $\varepsilon_2$ for $T \geq 0.2T_F$ while the results shown in Fig. 6 indicate that, the conductivity initially increases with increasing $\varepsilon_2$ and then decreases with $\varepsilon_2$ for $T = 0.01T_F$. So, a nonmonotonicity of the electrical conductivity of layer 1 as function of $\varepsilon_2$ is observed at low temperatures. Besides, Fig. 5 show that the transition temperature decrease with increasing $\varepsilon_2$. It reaches highest (smallest) value of $0.3T_F$ ($0.1T_F$) for $\varepsilon_2 = 1$ ($\varepsilon_2 = 22$). The results of the transition temperature are similar to the case that the charged impurities are located only in the first layer.

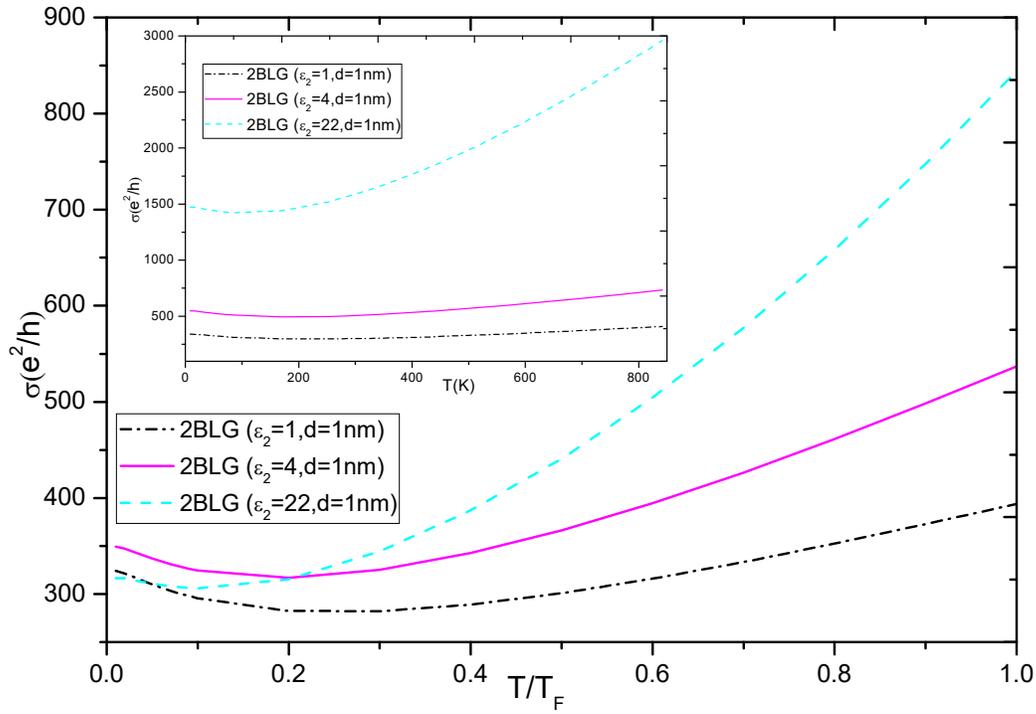

Fig. 5. The electrical conductivity $\sigma$ of layer 1 for 2BLG with $\varepsilon_3 = 12.53, d = 1 nm, n_{i,2} = 10 n_{i,1}$ as a function of $\frac{T}{T_F}$ for three values of $\varepsilon_2$. The insets show corresponding results for the case when the impurities are absent in layer 2, $n_{i,2} = 0$.

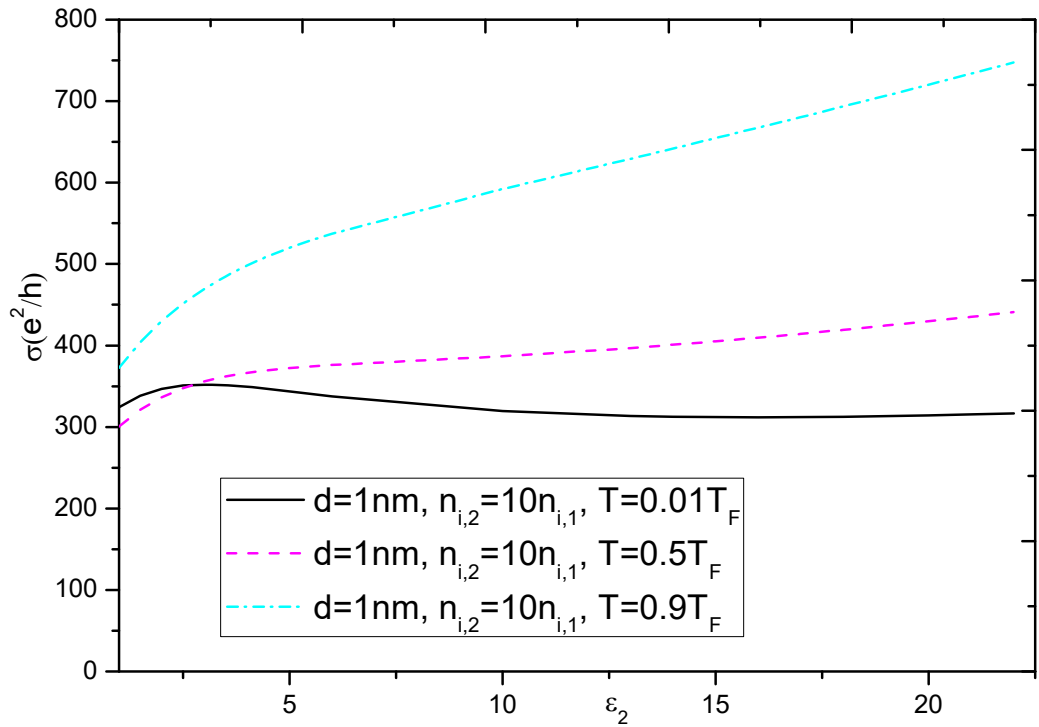

Fig. 6. The electrical conductivity of layer 1 of 2BLG with $\varepsilon_3 = 12.53, d = 1\ nm$ and $n_{i,2} = 10 n_{i,1}$ as a function of $\varepsilon_2$ for $T = 0.01 T_F$, $T = 0.5 T_F$ and $T = 0.9 T_F$.

### 3.4. The dependence of conductivity of the first layer on dielectric constant of substrate

In order to bring out the importance of substrate effects [41], we calculate the electrical conductivity σ for $\varepsilon_2 = 4$, $d = 1\ nm$ and $n_{i,2} = 10 n_{i,1}$ as a function of temperature for three different values of $\varepsilon_3$ and as a function of dielectric constant $\varepsilon_3$ for three different values of temperature $T$. The results shown in Fig. 7 and Fig. 8 indicate that σ increases with increasing $\varepsilon_3$ more rapidly compared to the case when the charged impurities are located only in the first layer [35]. So, we see that for the double layer system based on bilayer graphene, the substrate strongly reduce the Coulombic scattering only due to charged impurities located in layer II. Finally, Fig. 7 establish that the transition temperature is $0.2 T_F$ for all $\varepsilon_3$ which is similar to the case that the charged impurities are absent in layer 2.

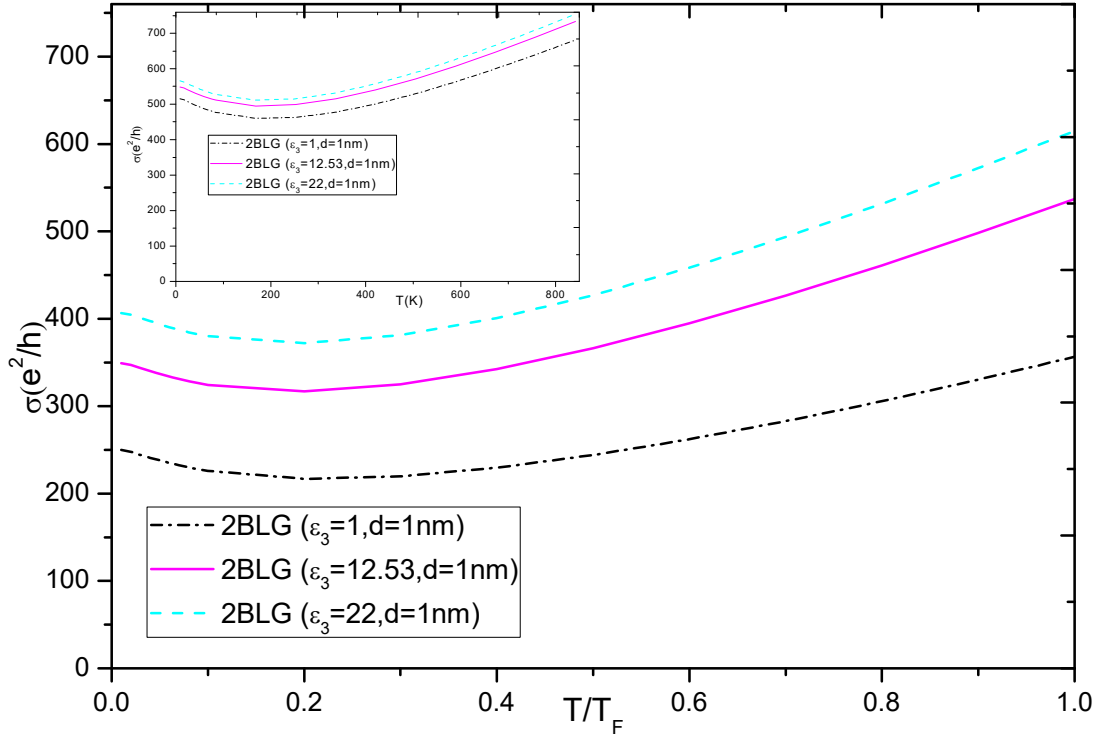

Fig. 7. The electrical conductivity σ of layer 1 for 2BLG with $\varepsilon_2 = 4.0$, $d = 1nm$, $n_{i,1} = 10 n_{i,2}$ as a function of $\frac{T}{T_F}$ for three values of $\varepsilon_3$. The insets show corresponding results for the case when the impurities are absent in layer 2, $n_{i,2} = 0$.

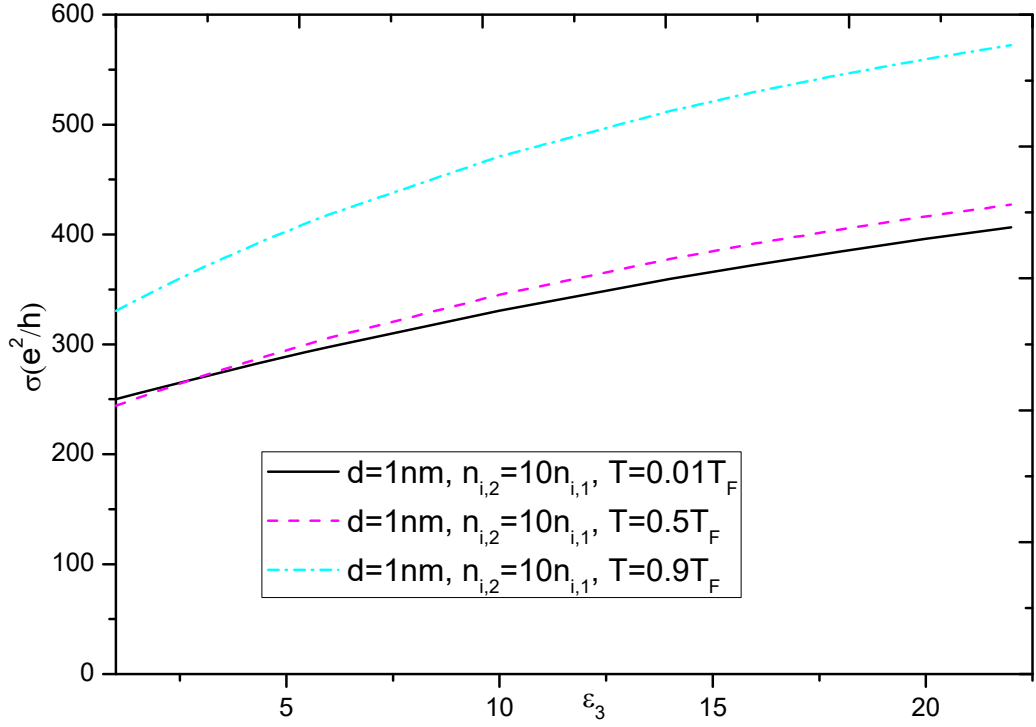

Fig. 8. The electrical conductivity of layer 1 of 2BLG with $\varepsilon_2 = 4, d = 1\ nm$ and $n_{i,2} = 10n_{i,1}$ as a function of $\varepsilon_3$ for $T = 0.01T_F$, $T = 0.5T_F$ and $T = 0.9T_F$.

### 3.5. The dependence of conductivity of the first layer on carrier concentration in layer II

Finally, we investigate behavior of the electrical conductivity of layer 1 with increasing carrier concentration $n_2$. We choose $n_2 = 2 \times 10^{12} - 4 \times 10^{12}$ cm$^{-2}$ [31] so that BLG ( layer 2) have a quadratic dispersion. We calculate the electrical conductivity σ for $\varepsilon_2 = 4, d = 1\ nm$, $n_1 = 2 \times 10^{12} cm^{-2}$, and $n_{i,2} = 10n_{i,1}$ as a function of temperature for three different values of $n_2$. For a comparison, we show in the inset for three different values of $n_2$, the temperature dependence of the conductivity of a 2BLG system that the impurities are absent in layer 2. The figure shows that the conductivity decreases with increasing carrier concentration $n_2$ for $T < 0.6T_F$ then remains almost constant for higher temperature. Besides, the conductivity depends on $n_2$ more strongly compared to the case which the charged impurities are present only in the layer 1. The transition temperature is found to be equal to $0.2T_F$ for all $n_2$

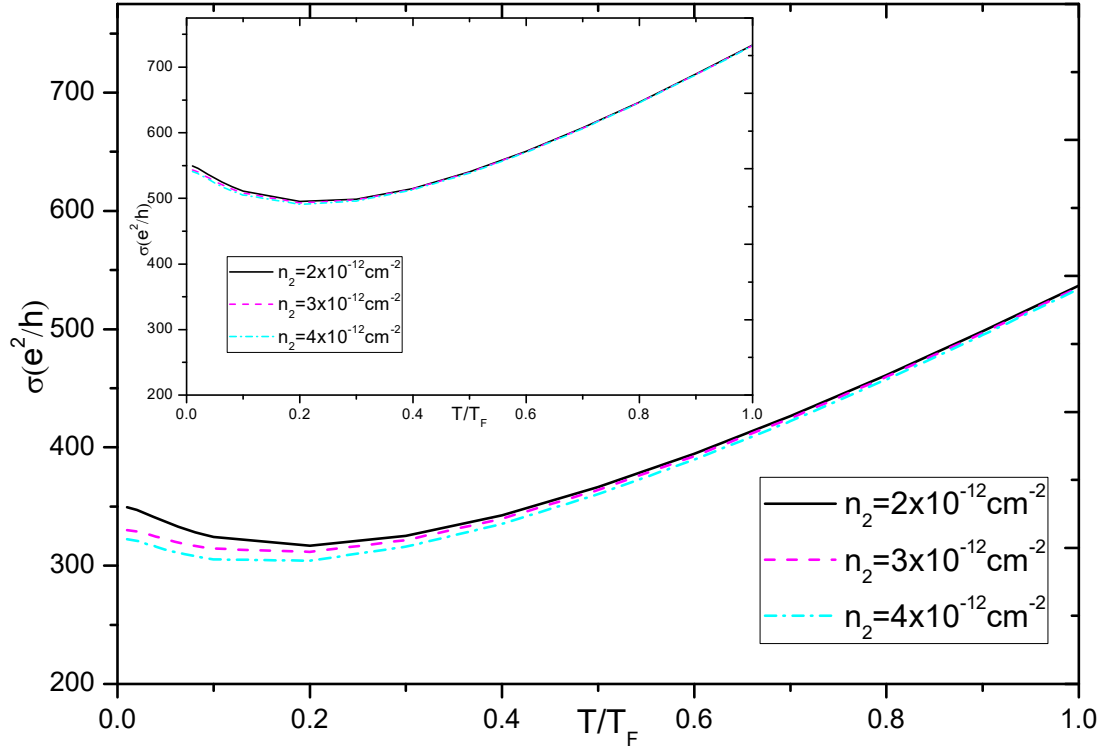

Fig. 9. The electrical conductivity $\sigma$ of layer 1 for 2BLG with $\varepsilon_2 = 4, d = 1nm, n_1 = 2 \times 10^{12} cm^{-2}, n_{i,1} = 10n_{i,2}$ as a function of $\frac{T}{T_F}$ for three values of $n_2$. The insets show corresponding results for the case when the impurities are absent in layer 2, $n_{i,2} = 0$.

## 4. Conclusion

In conclusion, we have considered a double layer BLG-BLG and have fully investigated the system parameter dependence of electrical conductivity of the first layer due to charged impurities located in both layers. Our results show that for small interlayer distances, $d = 1nm$, $\sigma$ decreases notably with increasing $n_{i,2}$ for $n_{i,2} \geq 7n_{i,1}$. However, for large interlayer spacing, $d \geq 4nm$, the effect of charged impurities in the second layer 2 is negligible. The temperature dependence of conductivity $\sigma$ are similar for all $d$. We also find that, when the effect of $n_{i,2}$ is apparent, the conductivity increases with increasing $\varepsilon_2$ for $T \geq 0.2T_F$ and shows a nonmonotonic behavior at low temperatures. Besides, our calculation indicates that σ increases (decreases) with increasing $\varepsilon_3(n_2)$ more rapidly compared to the case when the charged impurities are present only in the layer 1. It is seen that for the bilayer graphene double layers the substrate strongly reduce the Coulombic scattering only due to charged impurities located in layer II. The transition temperature doesn't depend on $d, n_{i2}, \varepsilon_3, n_2$, decreases with increasing $\varepsilon_2$ for $n_1 = n_2 = n = 2 \times 10^{12}$ cm$^{-2}$. Finally, σ increases with increasing carrier concentration $n_2$ for $T < 0.6T_F$ then remains almost constant for higher temperature.